\documentclass[aps,pre,twocolumn]{revtex4-1}

\usepackage{amsmath}
\usepackage{amsthm}

\usepackage{bbm,graphicx,bm,hyperref}   

\usepackage{dutchcal}

\makeatletter
\newcommand*{\defeq}{\mathrel{\rlap{%
                     \raisebox{0.3ex}{$\m@th\cdot$}}%
                     \raisebox{-0.3ex}{$\m@th\cdot$}}%
                     =}
\makeatother

\def\ii{{\rm i}}
\def\sx{\sigma^{\rm x}}
\def\sy{\sigma^{\rm y}}
\def\sz{\sigma^{\rm z}}
\def\tr#1{{\rm tr}(#1)}
\def\1{\mathbbm{1}}

\def\ket#1{{| #1 \rangle}}

\def\hx{h_{\rm x}}
\def\hz{h_{\rm z}}
\def\Sr{{\cal S}_r}

\def\tit#1{{\em #1},}

\begin{document}

\title{Momentum dependent quantum Ruelle-Pollicott resonances in translationally invariant many-body systems}

\author{Marko \v Znidari\v c}
\affiliation{Physics Department, Faculty of Mathematics and Physics, University of Ljubljana, 1000 Ljubljana, Slovenia}

\date{\today}

\begin{abstract}
  We study Ruelle-Pollicott resonances in translationally invariant quantum many-body lattice systems via spectra of a momentum-resolved operator propagator on infinite systems. Momentum dependence gives insight into the decay of correlation functions, showing that, depending on their symmetries, different correlation functions in general decay with different rates. Focusing on the kicked Ising model the spectrum seems to be typically composed of an annular random matrix like ring whose size we theoretically predict, and few isolated resonances. We identify several interesting regimes, including a mixing regime with a power-law decay of correlation functions. In that regime we also observe a huge difference in time-scales of different correlation functions due to an almost conserved operator. An exact expression for the singular values of the operator propagator is conjectured, showing that it becomes singular at a special point.
\end{abstract}




\maketitle

\section{Introduction}

One of the outstanding questions of theoretical physics is understanding different possible dynamical properties that systems can display. In classical mechanics the theory of dynamical systems~\cite{ott,gaspard} provides a theoretical framework that is well understood for few particle systems. Most of the objects we can measure in physics are either simple expectations of observables, or their correlation functions. Correlation functions are not just experimentally important but allow also to distinguish different dynamical regimes, for instance, an exponential decay will typically occur in chaotic systems, while no decay implies that the system is at most ergodic but not mixing. One way to approach such questions is using the so-called Ruelle-Pollicott (RP) resonances~\cite{Pollicott85,Ruelle86}. The formalism is theoretically appealing -- one can treat classical and quantum systems on essentially equal footing -- and it provides decay rates with which correlation functions decay, and also, as we shall explain, literally forces one to ask ``the right questions''.

While our paper focuses on much less explored quantum RP resonances, let us start with classical systems. Dynamics in phase space can be studied with a Frobenius-Perron (FP) propagator $U$, which is a linear operator that propagates densities on phase space. Equivalently, one could also study its adjoint called a Koopman operator that instead propagates observables~\cite{Braun,gaspard}. The FP operator is unitary on the space of $L^2$ functions with the spectrum therefore being on a unit circle $|z|=1$. However, analytically continuing the resolvent $1/(z-U)$ to $|z|<1$ one can identify singularities within a unit circle. They will in particular determine the asymptotic decay of correlation functions of smooth initial densities as $\asymp |\lambda_1|^t$, where $\lambda_1$ is the resonance with the largest modulus. In an integrable system one expects $|\lambda_1| \to 1$, while in a chaotic system one should have an isolated $|\lambda_1|<1$ which are called RP resonances~\cite{Pollicott85,Ruelle86}. Note that, despite starting with a unitary $U$ on the space $L^2$, we are effectively dealing with a non-unitary operator (on a space that is not $L^2$).

As is often the case tough, theoretical elegance comes at a price. It is not clear how to in practice do the required ``analytical continuation''. Mathematics is heavy, and even in exactly solvable classical cases like the cat map~\cite{antoniou97}, the Bernoulli map or the baker map~\cite{Hasegawa92,Tasaki93} (see Ref.~\cite{gaspard95} for differential equations), one often rather starts by writing the spectral decomposition of $U$ on an appropriate system-dependent functional space. Thereby, the crux of the problem shifts to the question of what is the appropriate functional space.

A more practical approach, and the only viable option in numerical calculations of RP resonances, is to make a non-unitary operator $U(\epsilon)$ out of a unitary $U$ by adding some ``dissipation'' and then identifying RP resonances as isolated eigenvalues in the limit of zero dissipation strength $\epsilon$. Dissipation is often added in the form of a noise or doing some coarse graining. Such a noisy or a coarse graining procedure adds a practical appeal to an otherwise rather mathematical formulation of RP resonances. In real life we are never really in the unitary limit -- the system of interest is always invariably coupled to some extra degrees of freedom, and/or, in a many-body (quantum) system we never have access to all observables and therefore full unitary evolution. If we study a system numerically we also necessarily have numerical errors, meaning that we are again studying a noisy non-unitary system. From a practical point of view an exact unitarity is therefore just a mirage. Non-unitary prescription to get the RP resonances is exactly what one anyway has to deal with. Let us finish this introduction with a few words about the spectral decomposition of such non-unitary linear operators.

As we have mentioned, for some classical solvable maps~\cite{Hasegawa92,Tasaki93,antoniou97} one can explicitly calculate the spectral decomposition. From those solutions one sees that the eigenvectors are rather singular objects -- they involve Dirac delta functions and their derivatives -- and are therefore ``too wild'' to be in $L^2$. They are not really smooth densities but rather functionals. The space on which these functionals are allowed to act also excludes for instance Dirac delta functions, meaning that the intricacies of mathematical formalism actually forces us to avoid asking meaningless questions, such as about individual trajectories in a chaotic unstable system, and instead focus on the evolution of smooth densities (in quantum mechanics on the evolution of density operators rather than of pure states). One can also reach the same conclusion about the singular behavior of spectral decomposition in a more intuitive way: a common scenario of chaos in smooth classical systems is the so-called stretch-and-fold mechanism, see e.g. Refs.~\cite{chaosbook,ott}, due to stable and unstable manifolds. As one evolves an initial smooth density $\rho(0)$ it is stretched along the unstable manifold. Therefore, $\rho(t)$ becomes smoother along the unstable manifold, and at the same time less smooth along the stable manifold. At long times $\rho(t)$ will therefore become localized around the unstable manifold. With the spectral decomposition in mind this suggests that the right eigenvectors $\ket{R_j}$ will be localized along the unstable manifold, while the left ones $\ket{L_j}$ will be localized along the stable manifold. Because in a chaotic system those manifolds have a fractal structure with details on an arbitrarily small scale, one expect similar singular (fractal) behavior also in eigenvectors. That is fine though, again, in accordance with chaotic systems having a positive rate of generating information (as measured e.g. by dynamical entropies~\cite{ott,gaspard}), this high complexity is apparently hidden in eigenvectors, which, though, on themselves do not have much ``meaning'' -- only their action on smooth initial densities is what is physically observable. Physical information about relaxation rates toward equilibrium is on the other hand simply encoded in eigenvalues $\lambda_j$~\cite{foot5}. This splitting between information contained in eigenvectors and eigenvalues is less clear if one instead looks on a unitary propagator $U$.

Going to quantum physics, all of the above comments and ideas essentially carry over by starting with a unitary propagator of states $U$, or, as we will do, considering the adjoint Heisenberg evolution of operators instead.

\subsection{Previous work}

Exact results on RP resonances are possible for classical hyperbolic maps, like the baker and the Bernoulli map~\cite{Hasegawa92,Tasaki93}, and the cat map~\cite{antoniou97}. The standard map has been studied in Refs.~\cite{fishman00,venegeroles08,mata23}, while the kicked top, including a regime where classical dynamics is mixed, in Ref.~\cite{weber}. A detailed analysis of the multibaker map is in Ref.~\cite{gaspard92,gaspard}, while coupled baker maps have been studied in Refs.~\cite{arul94,Haake03}. Rather than working with a propagator one can approach everything from the point of view of spectral analysis of correlation functions, extending frequencies to the complex domain~\cite{gaspard}. Thereby one extracts RP resonances directly from correlations, see Ref.~\cite{martin} for an example. More rigorous treatment and a comparison between classical and quantum cases with noise for maps on a torus can be found in Ref.~\cite{Nonnenmacher}. Another technical trick of how to obtain RP resonances is using periodic orbits with trace formulas and zeta functions~\cite{chaosbook,sridhar,altshuler,leboeuf}. It has been observed that sometimes there might be issues with convergence of RP resonances for particular choices of coarse-graining~\cite{blank98,shepelyansky,shudo21} (which could perhaps be related to strong non-normality of matrices).

In single-particle quantum chaos~\cite{Haake} spectral criteria of quantum chaos, e.g., nearest-neighbor level spacing, proved to be useful. In many-body systems the mean level spacing is exponentially small and therefore in some situations spectral quantum chaos criteria at the smallest energy scales may be inappropriate. The RP formalism instead is a setting that can be used. The quantized perturbed cat map with noisy shift being described by a Kraus map has been studied in Ref.~\cite{mata0304}. More recently it has been proposed~\cite{Mori} to use Lindblad dissipation, calculating the Lindbadian gap in an appropriate limit $\lim_{\epsilon \to 0}\lim_{L \to \infty}$ (note that this is the difficult order of limits to evaluate as one usually has to get the gap in the thermodynamic limit at a finite $\epsilon$), see also results in Ref.~\cite{sa} that can be used to that end. Last two works study many-body quantum systems, in particular Ref.~\cite{Mori} the kicked Ising (KI) model whose RP resonances were studied for the first time already more than two decades ago in Refs~\cite{Prosen,Prosen07}.

\subsection{New results}

We approach RP resonances by studying the spectrum of a propagator of operators on an infinite system. Resolving the (quasi)momentum dependence and truncating it down to operators with support on $r$ consecutive sites, we can discuss the decay of any correlation function. To illustrate our results we are going to study the same kicked Ising model and use essentially the same truncation method used in Ref.~\cite{Prosen}, but extend it from the zero momentum subspace studied there. Doing that we will find many new and interesting results. As we work directly in the thermodynamic limit, i.e., study the evolution of operators in an infinite system, there is only one limit we need to take, namely, the operator support $r$ to infinity, instead of typically two for other types of coarse-graining or noise approaches to RP resonances. We show that to properly assess the dynamical properties of the system it pays to take into account the momentum dependence of RP resonances because different observables in general decay with different rates. The spectrum of eigenvalues in the KI model consists of a random-matrix like bulk of an annular shape whose radii we analytically predict, and few isolated resonances. Those indicate rich physics: besides identifying expected and known chaotic regime, we also find a mixing non-chaotic regime with a power-law decay of correlations due to a branch cut. This regime is rather interesting because, despite being away from any integrable point, different observables decay on vastly different timescales. For instance, for a particular choice of parameters magnetization correlations start to decay on a $\sim 10^4$ times longer time scale than for instance the staggered magnetization or magnetization current correlations. The origin of such disparate timescales is due to an almost conserved operator giving rise to a prethermalization plateau. Using an analytical expression for singular values of the truncated propagator we also conjecture a lower bound on RP resonances that depends only on the parameter $\tau$. Singular values also highlight the speciality of KI parameters when $\tau$ is an odd multiple of $\pi/4$, for which some singular values become $0$ (this point includes an even more special dual unitary KI model~\cite{DU} if one in addition sets $\hx=1$).

\subsection{Kicked Ising model}

RP resonances in the KI model have been already studied in Refs.~\cite{Prosen,Prosen07}, and more recently in Ref.~\cite{Mori}. For numerical demonstrations we will take the same model, one reason being simply to have a reference point with already known results. The model can be defined in terms of a time periodic Hamiltonian,
\begin{eqnarray}
  H&=&H_{\rm z}+ H_{\rm x}\sum_{n=-\infty}^{\infty}\tau \delta(t-n),\\
  H_{\rm x}&=&-\hx \sum_j \sx_j,\nonumber \\
  H_{\rm z}&=&-\sum_j \sz_j \sz_{j+1}-\hz\sum_j \sz_j. \nonumber
  \label{eq:KI}
\end{eqnarray}
Equivalently, we can directly integrate its evolution over one period of duration $\tau$, writing a one-step Floquet propagator $U$ as
\begin{eqnarray}
  U&=& U_{\rm x} U_{\rm z},\\
  U_{\rm x}&=&{\rm e}^{-\ii H_{\rm x}\tau}=\prod_j {\rm e}^{\ii \tau \hx \sx_j},\nonumber \\
  U_{\rm z}&=&{\rm e}^{-\ii H_{\rm z} \tau}={\rm e}^{\ii \tau \sum_j \sz_j \sz_{j+1}}\prod_j {\rm e}^{\ii \tau \hz \sz_j}. \nonumber
  \label{eq:KIU}
\end{eqnarray}
There are three parameters $\tau$, $\hx$, and $\hz$ (we use the same parameterization as in Ref.~\cite{Mori}, see Appendix~\ref{app:Par} for an alternative used in Ref.~\cite{Prosen}). For generic parameters the model is nonintegrable without any local conserved quantities. We measure time in units of $\tau$, i.e., our $t$ takes integer values with the propagator being $U^t$.

The KI model is integrable at special values of parameters. Known integrable points are (i) $\tau \hx=0$ (or an integer multiple of $\pi$) for which the Hamiltonian is trivially diagonal in the computational basis, (ii) $\tau\hz=0$ (or an integer multiple of $\pi$), i.e., transversal field, for which it is integrable~\cite{prosenPTPS}, similar to integrability of the autonomous transversal Ising model, (iii) integrable points when the strength of the field $\gamma$ defined in Eq.(\ref{eq:gamma}) is a multiple of $\pi/2$~\cite{Prosen07}. This is achieved when either $\tau \hx$ or $\tau \hz$ is an odd multiple of $\pi/2$, or when both $\tau \hx$ and $\tau \hz$ are integer multiples of $\pi$ (for our choice of $\hx=0.9$ and $\hz=0.8$ two integrable points with the smallest $\tau$ are $\tau \approx 1.75$ and $\tau \approx 1.96$, shown with black points in Fig.~\ref{fig:odtau}). New trivial integrable point (iv) is when $\tau=\pi/2$ (or an odd multiple of $\pi/2$; the leftmost black point in Fig.~\ref{fig:odtau}) and arbitrary $\hx$ and $\hz$, for which the model is noninteracting because ${\rm e}^{\ii \sz_1 \sz_2 \pi/2}=\ii \sz_1 \sz_2$, and therefore the whole interacting part is trivial $\prod_j {\rm e}^{\ii \tau \sz_j \sz_{j+1}}=(\ii)^L \1$ (for PBC). Some other non-integrale special points will be mentioned in Sec.~\ref{sec:M}.

The propagator $U$ has several appealing properties (which was one of the reasons it was chosen in Ref.~\cite{Prosen}): it can be viewed as a quantum circuit made of commuting 1-qubit $\sz$ gates, then commuting 2-qubit $\sz_j\sz_{j+1}$ gates, followed finally by again commuting 1-qubit $\sx$ gates. The fact that it has a single layer of commuting 2-qubit gates means that the operator support can increase by at most $2$ sites in a single step. For instance, propagating a translationally invariant (TI) operator with support on $r=3$ sites will after one step result in operators with support on at most $r'=5$ sites, e.g.,
\begin{equation}
 \sum_j U^\dagger \sx_{j-1}\sz_j\sy_{j+1} U= \sum_j \sum_{\alpha,\beta,\gamma,\delta,\nu} \alpha_{j-2}\beta_{j-1}\gamma_j\delta_{j+1}\nu_{j+2},\nonumber
  \end{equation}
with single-site operators $\alpha,\beta,\gamma,\delta,\nu$. The idea of the truncated operator propagator $M$ is~\cite{Prosen} to limit oneself to operators with support smaller than some $r$, and then study possible isolated eigenvalues that are stable as one increases $r \to \infty$ (RP resonances).

The main physical quantity that we study will be autocorrelation functions $C(t)$ of different extensive observables $A=\sum_j a_j {\rm e}^{\ii k j}$,
\begin{equation}
  C(t)=\frac{1}{L}\left(\langle A(t)A \rangle-\langle A(t)\rangle \langle A \rangle \right),
  \label{eq:C}
\end{equation}
where the prefactor $1/L$ ensures the asymptotic independence on the system size $L$ in a finite system, and $k$ is the (quasi)momentum. The expectation value is an infinite temperature average, $\langle \bullet \rangle:=\frac{1}{2^L}\tr{\bullet}$, however in practice, because our system sizes are large enough, it typically suffices to evaluate it in a single random initial state $\psi$, $\langle \bullet \rangle=\langle \psi |\bullet|\psi \rangle$. Namely, fluctuations of $C(t)$ in a single random state are of the order $\sim 1/2^{L/2}$ (e.g., $\approx 10^{-5}$ at $L=32$). We will also briefly consider local observables.

Autocorrelation function $C(t)$ will be calculated in three different ways: (i) evaluating Eq.(\ref{eq:C}) in a finite system with periodic boundary conditions ($L\le 33$) by an exactly evolving random initial state, (ii) approximating it by a truncated operator propagator on an infinite system, and (iii) calculating the leading RP resonance of the truncated propagator, giving the asymptotic decay $\asymp |\lambda_1|^t$.

\section{Momentum-dependent spectrum of Ruelle-Pollicott resonances}

Many-body lattice systems with translational invariance have a conserved quasi-momentum $k$ (we will simply call it momentum). Compared to single-particle systems that do not have any spatial degree of freedom one should therefore expect a one-parameter family of RP resonances labeled by that momentum (and any other additional symmetries that one might have). In this section we are going to describe how to write a truncated operator propagator $M(k)$ in the subspace of momentum $k$. The procedure is a generalization of the truncation in the zero momentum sector used in Ref.~\cite{Prosen}.

\subsection{Truncated propagator of translated local operators}

Let us choose the local operator basis $\ell$ as Pauli matrices plus the identity,
\begin{equation}
  \ell=\mathcal{p} \cup \{ \1 \},\quad   \mathcal{p}=\{ \sx,\sy,\sz\},
\end{equation}
in short, the Pauli basis. The operator basis on $r$ sites is then formed by a direct product of operators from the Pauli basis on different sites $j$. Let us classify local operators according to their maximal support size $r$. The set of operators $\Sr$ is made of all operators that start with a non-identity operator on site $1$, i.e., one in the set $\mathcal{p}_1=\{ \sx_1,\sy_1,\sz_1\}$, and have the last non-identity operator at most at the site $r$. More precisely, $\Sr$ is a set of all operators
\begin{equation}
  \Sr=\{ a=b_1 b_2 \cdots b_r \vert (b_1 \in \mathcal{p})\, \land \, (b_{j=2,\ldots,r} \in \mathcal{l}) \}.
\end{equation}
As an example
\begin{equation}
  {\cal S}_1=\{\sx_1,\sy_1,\sz_1\},\qquad {\cal S}_2=\mathcal{p}_1\otimes \mathcal{l}_2.
\end{equation}
The number of elements $N$ in ${\cal S}_r$ is,
\begin{equation}
  N={\rm dim}(\Sr)=3\cdot 4^{r-1}.
  \label{eq:N}
  \end{equation}

The translation (super)operator by one site is denoted by $T$, and acts like e.g. $T(\sx_1\sz_2)=\sx_2 \sz_3$. The operator basis can be organized according to the momentum $k$ that labels eigenvalues ${\rm e}^{\ii k}$ of $T$. In a finite system $k$ takes a discrete set of values, while in the infinite lattice it is a continuous variable $k \in [0,2\pi)$. The $k$-resolved basis with support on at most $r$ sites is therefore composed of operators $A$ that are a translational sum of a local operator $a$,
  \begin{equation}
    A=\sum_j {\rm e}^{-\ii k j} T^j(a),\quad a \in \Sr.
    \label{eq:A}
  \end{equation}

  Because $H$ commutes with $T$, it has a block structure and one can treat dynamics in each $k$-subspace separately. Unitary Floquet one-step propagator $U$ induces evolution on the space of operators as $O'=U^\dagger O U$. Such evolution is unitary on the whole infinite dimensional operator space.
  However, if one truncates the operator space to translational sums of operators from $\Sr$, like those in Eq.~(\ref{eq:A}), one ends up with a nonunitary truncated propagator $M(k)$. Eigenvalues of $M(k)$ will be our RP resonances in which we are interested. Note that $M(k)$ directly acts on an infinite system and so the only remaining parameter is the support size $r$ that also determines the size $N$ (\ref{eq:N}) of the matrix $M(k)$.

  For our KI system (\ref{eq:KIU}) constructing $M(k)$ is easy. As we explained, one application of $U$ can increase the support of $A$ only by one site at each edge of $a$. Therefore, to get the matrix element $[M(k)]_{A,A'}$ between two translational operators $A$ and $A'$, defined in terms of local $a$ and $a'$ from $\Sr$ (\ref{eq:A}), respectively, it suffices to act with $U$ on an operator supported on $r+2$ sites. That is, starting with $a' \in \Sr$ supported on sites $1,\ldots,r$ we extend it to $r+2$ sites by concatenating it with identities on each end to accommodate for a possible spreading of $a'$ at each end by at most one site, and then transform it with $U$. The resulting transformed $a'$ can be expanded in terms of operators $b$ on $r+2$ sites, 
  \begin{equation}
    U^\dagger (\1_0\, a'\, \1_{r+1}) U=\sum_b c_b(a')\, b,
    \label{eq:ca}
  \end{equation}
where the sum on the RHS is over all possible operators $b$ on $r+2$ sites and $c_b(a')$ are corresponding expansion coefficients (which depend on $a'$). The matrix element of the truncated propagator can then be simply expressed in terms of expansion coefficients as,
  \begin{equation}
    [M(k)]_{A,A'}= c_a(a')+{\rm e}^{-\ii k} c_{T^{-1}a}(a')+{\rm e}^{\ii k} c_{Ta}(a').
    \label{eq:M}
    \end{equation}
Compared to other possible ways of obtaining RP resonances, like using noise or some coarse-graining, this truncation procedure has several appealing features~\cite{Prosen}. It is physically motivated as one is usually interested in local operators, and, because one directly works in the thermodynamic limit, one has to take only one limit, that is $r \to \infty$, instead of usually two like the noise strength $\epsilon \to 0$ as well as the system size $L \to \infty$. The matrix $M(k)$ is actually equal to a truncation of a unitary propagator in the $k$ sector of a finite system with PBC, provided $L \ge 2r-1$ (i.e., taking a finite system with PBC one selects in a given momentum block only operators with support on $\le r$ sites). Therefore one can consider our infinite system size $M(k)$ as a specific operator-support selective way of truncating an appropriate unitary matrix. Let us mention at this point that within a random matrix theory one has studied truncations of Haar random unitary matrices, and they result in a spectrum inside a circle whose radius depends on the fraction of retained vectors~\cite{karol00}. 

  When numerically dealing with $M(k)$ we never really write out the full matrix (its size $N$ gets too large for large $r$, e.g., $N\approx 5\cdot 10^7$ for $r=13$), but rather generate it's action on the fly by evaluating Eq.(\ref{eq:ca}). This is done by applying each gate of the quantum circuit representation of $U$, see Appendix~\ref{app:matrix}.

Due to translational symmetry one also has that $M(-k)$ has the same spectrum as $M(k)$ and therefore one can limit oneself to $k \in [0,\pi)$, rather than $k \in [0,2\pi)$ (the spectrum at $2\pi-k$ is the same as at $k$). In addition to that, due to a spatial reflection symmetry (reversing the order of operators, i.e., in a finite system corresponding to sites mapping $j \to L-j$), which the KI Hamiltonian (\ref{eq:KI}) has, the $k=0$ sector splits into two independent parity subsectors corresponding to eigenvalues $\pm 1$ of the reflection operator. Namely, $N$ elements of the local basis $\Sr$ can be split into two sets according to whether the reflected operator $Ra$ is the same as $a$, where the reflection of the local $a$ is defined by
      \begin{equation}
        R(a_1 a_2\cdots a_p)=a_p a_{p-1} \cdots a_1,
      \end{equation}
with $p$ being the index of the rightmost operator that is not equal to identity, i.e., $a_p \in \mathcal{p}$ but $a_{p+1,\ldots,r}=\1$. That is $\Sr=\Sr^{\rm pairs} \cup \Sr^{\rm single}$, where $\Sr^{\rm single}=\{a \vert (a \in \Sr)\, \land\, (Ra=a) \}$, and $\Sr^{\rm pairs}=\{a \vert (a \in \Sr)\, \land\, (Ra\neq a) \}$. Instead of using the basis given by the elements of $\Sr$, one takes an odd basis $\{ (a-Ra)/\sqrt{2} \vert a \in \Sr^{\rm pairs}\}$, and an even basis $\{ a \vert a \in \Sr^{\rm single} \} \cup \{ (a+Ra)/\sqrt{2} \vert a \in \Sr^{\rm pairs} \}$ (where each pair is of course accounted for only once and not twice). The corresponding even and odd sector matrices $M(k=0^+)$ and $M(k=0^-)$ are then constructed as in Eqs.(\ref{eq:ca}) and (\ref{eq:M}) with the only difference being that $a'$ and $a$ run over elements of the even or the odd basis, instead of over the full $\Sr$.

 As we shall see, studying the momentum dependent RP resonances will give us much more information than just limiting ourselves to $k=0^+$ (a sector to which e.g. total magnetization belongs). In particular, different dynamical regimes, like chaotic, mixing, and integrable, will be easier to identify. Note also that some important frequently studied operators are not from the $k=0^+$ sector, an example being the magnetization current (i.e., charge current in fermionic language) operator $J$ (the KI model of course does not conserve the magnetization)
      \begin{equation}
        J=\sum_j (\sx_j \sy_{j+1}-\sy_{j}\sx_{j+1}),
        \label{eq:J}
      \end{equation}
which is from the $k=0^-$ sector. Another is the staggered magnetization, denoted by $sZ$, (also called the imbalance)
      \begin{equation}
        sZ=\sum_j (-1)^j \sz_j,
        \label{eq:sZ}
      \end{equation}
 which is from the sector with $k=\pi$. Also, by combining different $k$ sectors (Fourier transformation) one can as well treat non-extensive operators, for instance, strictly local ones.

\subsection{Some properties of $M(k)$}
\label{sec:M}

Let us mention at this point that the properties of matrix $M(k)$ are rather interesting and warrant future studies. As already noted in Ref.~\cite{Prosen}, when studying the $k=0$ case, $M$ has a self-similar structure (which is also reflected in a self-similar structure of eigenvectors). Here we just list few numerical observations that will help us understand the behavior of RP resonances and the decay of correlations.

We start with singular values of $M(k)$ that turn out to be independent of $k$ as well as of $\hx$ and $\hz$. They depend only on $\tau$. In fact, there are only 3 different singular values having a simple analytical form,
\begin{equation}
  s_1=1,\quad s_2=|\cos{(2\tau)}|,\quad s_3=\cos^2{(2\tau)},
  \label{eq:s}
\end{equation}
with multiplicities of respectively
\begin{equation}
5\cdot 4^{r-2},\quad 2\cdot 4^{r-2},\quad 5\cdot 4^{r-2}.
\end{equation}
Independence of singular values on the parameters of single qubit gates is easy to see. Namely, the unitary propagator (\ref{eq:KIU}) is a product of single qubit gates and two qubit gates. Because the truncation we use is a truncation in support of operators, and single qubit gates do not change the support, the truncated propagator can be written as a product of two matrices,
\begin{equation}
  M=O M_{\rm zz},
  \label{eq:OM}
\end{equation}
where $M_{\rm zz}$ is non-unitary and corresponds to transformation by all two qubit gates $U_{\rm zz}={\rm e}^{\ii \tau \sum_j \sz_j \sz_{j=1}}$, while $O$ is an orthogonal matrix and describes the action of the remaining single qubit gates. Eq.(\ref{eq:OM}) immediately means that $M$ and $M_{\rm zz}$ have the same singular values (though different eigenvalues!). Therefore, the singular values of the full $M$ depend only on $\tau$. Their exact functional form should presumably follow from the simple form of $M_{\rm zz}$.

Moving to eigenvalues, as we shall see later (Fig.~\ref{fig:045spek}), the bulk of the spectrum of eigenvalues of $M(k)$ will consist of an annular cloud of eigenvalues. Such an annular region of eigenvalues is for instance obtained~\cite{foot3} in a singular value decomposition motivated random matrix ensemble of form $U\,s\,V$, where $U$ and $V$ are Haar random unitary while $s$ is a fixed real diagonal matrix (of e.g. singular values)~\cite{zee01,ring}. A nice single-ring theorem also says~\cite{zee01} that one can express the inner radius $r_{\rm in}$ and the outer one $r_{\rm out}$ of the annulus in terms of averages over singular values $s_j$,
\begin{equation}
  r_{\rm in}=1/\sqrt{\langle s_j^{-2} \rangle},\quad r_{\rm out}=\sqrt{\langle s_j^2 \rangle}.
  \label{eq:r}
\end{equation}
In our case $U$ and $V$ are of course not really Haar random unitaries -- $M(k)$ is a specific fixed matrix -- nevertheless, Eq.(\ref{eq:r}) seems to give a rather good description (Fig.~\ref{fig:045spek}). Using the above singular values (\ref{eq:s}) and their multiplicities we can therefore estimate that the ring will have radii
\begin{eqnarray}
  r_{\rm in}&=&\sqrt{\frac{12}{5+\frac{2}{\cos^2{(2\tau)}}+\frac{5}{\cos^4{(2\tau)}}}},\nonumber \\
  r_{\rm out}&=&\sqrt{\frac{5+2\cos^2{(2\tau)}+5\cos^4{(2\tau)}}{12}}.
  \label{eq:myr}
\end{eqnarray}
Because the RP resonances need to be isolated from this RMT cloud of eigenvalues we conjecture a lower bound on any RP resonance $\lambda_j(k,\tau,\hx,\hz)$ as
\begin{equation}
  |\lambda_j| > r_{\rm out},
  \label{eq:bound}
\end{equation}
where $r_{\rm out}$ is given by Eq.(\ref{eq:myr}). The bound in particular limits the smallest possible leading RP resonance, regardless of parameters, to be $|\lambda_1| > {\rm min}(r_{\rm out})=\sqrt{5/12}\approx 0.64$ (see also brown curve in Fig.~\ref{fig:odtau}). In the cases we tested the bound (\ref{eq:bound}) is always satisfied~\cite{foot2}. 

From singular values (\ref{eq:s}) we see that $\tau=\pi/4$ is a special point (as well as any odd multiple of $\pi/4$). This is the point where the interaction gate of the KI becomes a signed permutation matrix (Appendix~\ref{app:matrix}), and the truncated propagator becomes singular -- there are only $5\cdot 4^{r-2}$ nonzero singular values that are all $1$ with the rest being zero. This is also the number of nonzero eigenvalues of $M(k)$ at $\tau=\pi/4$. If one in addition sets $\hx=1$, i.e., a one parameter family of KI models with $\tau=\pi/4$ (or an odd multiple of $\pi/4$) and $\hx=1$, one has an even more special situation with $U$ being dual unitary~\cite{DU}. For the dual unitary case we note that while the number of nonzero singular values stays the same $5\cdot 4^{r-2}$, the number of nonzero eigenvalues of $M(k)$ further decreases to just $2^{r-1}$. They are all independent of $k$ (in-line with a spatial delta function dependence of correlations which are for dual unitary circuits nonzero only on a light-cone), though they do depend on $\hz$. If one also sets $\hz=1$ the KI model circuit is dual unitary as well as Clifford, and all eigenvalues of $M(k)$ become $0$. These interesting properties suggest further study of the case $\tau=\pi/4$.

\begin{figure}[t!]
  \centerline{\includegraphics[width=3.in]{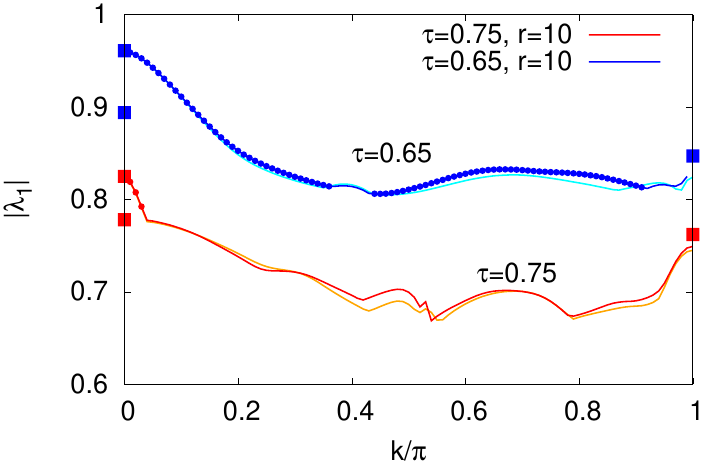}}
\caption{Dependence of the leading RP eigenvalue on $k$, all for $h_x=0.9$, $h_z=0.8$. Full squares show the values in the limit $r \to \infty$ for $k=0$ (upper square for $0^+$, lower for $0^-$) and $k=\pi$. Lighter colored curves (cyan for $\tau=0.65$ and orange for $\tau=0.75$) show results for $r=9$ indicating how fast $|\lambda_1|$ converges with $r$. Small points along the curves indicate values of $k$ where $\lambda_1$ is real (for $\tau=0.65$ these are most of $k$, for $\tau=0.75$ only very small ones before the first kink).}
\label{fig:odk}
\end{figure}
Finally, possible eigenvalues $\lambda=1$ of $M(k)$ at a finite truncation $r$ would indicate an exact conserved local operator supported on at most $r$ sites. Because our model is in general chaotic without any conservation laws there are no such eigenvalues except at mentioned special integrable points. Note that the translation operator $T$ can not be written as an exponential of a local generator that would be called a (quasi)momentum operator. While $T$ can be written as a product of nearest-neighbor swap gates $T_{j,j+1}=\exp{[-\ii \frac{\pi}{4}(\sx_j \sx_{j+1}+\sy_j\sy_{j+1}+\sz_j\sz_{j+1}-\1)]}$, trying to write it as $T={\rm e}^{\ii\, \hat{p}}$, the generator $\hat{p}$ is not local. Therefore, we don't have any eigenvalue $1$ (or close to $1$) of $M(k=0)$ for any finite $r$ that would correspond to the (quasi)momentum $\hat{p}$. In other words, for any finite $r$ the (quasi)momentum operator $\hat{p}$ is not (even close to) conserved. This makes lattice models with our truncation a bit different than continuous systems in an appropriate basis where presumably $\hat{p} \sim -\ii \partial_x$ would be included.

\subsection{Kicked Ising resonances}

Let us now numerically calculate~\cite{foot1} the leading eigenvalues of the matrix $M(k)$ (\ref{eq:M}) for the KI model (\ref{eq:KI}). We pick parameters similar to those used in Ref.~\cite{Mori}, namely $\hx=0.9$ and $\hz=0.8$, and two values of $\tau=0.65$ and $\tau=0.75$. In Fig.~\ref{fig:odk} we show the $k$-dependence of the eigenvalue $\lambda_1$ with the largest modulus. We can see several interesting features. First, the largest eigenvalue comes from the $k=0^+$ sector. The $k$ dependence is expectedly nontrivial, with several kinks visible at values where one has a collision of eigenvalues.
\begin{figure}[t!]
  \centerline{\includegraphics[width=3.in]{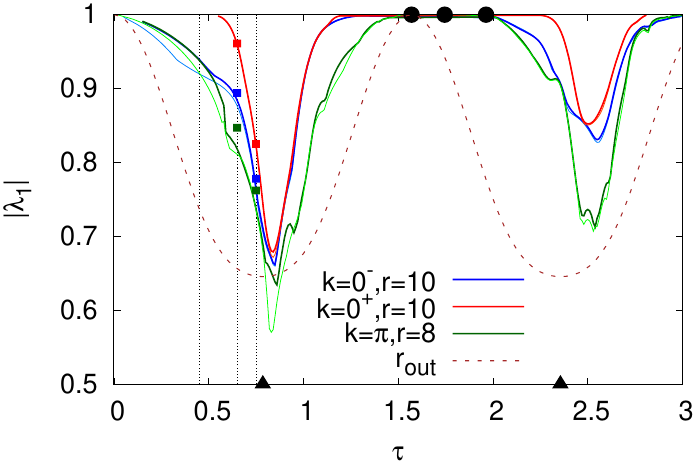}}
\caption{Dependence of the leading RP resonance on $\tau$ for three selected momentum sectors. All is for $h_x=0.9$, $h_z=0.8$. Lighter colored curves are for smaller $r$ in order to indicate convergence with $r$; light blue for $r=8$ and $k=0^-$, and light green for $r=7$ and $k=\pi$. Three vertical dashed lines are at $\tau=0.45,0.65,0.75$, which we study in detail in the following. Back circles at the top figure edge mark integrable values of $\tau$, while two black triangles at the bottom edge mark $\tau=\pi/4,3\pi/4$ at which $M(k)$ becomes singular. Brown dashed curve is the bound (\ref{eq:bound}). Filled squares at $\tau=0.65$ and $\tau=0.75$ are resonance values for $r \to \infty$.}
\label{fig:odtau}
\end{figure}

Even more interesting and informative is the dependence of the leading resonance on $\tau$. In Fig.~\ref{fig:odtau} we focus on three different momentum sectors, $k=0^{\pm}$ and $k=\pi$, in which the leading RP resonance is expected to give the decay rate of the autocorrelation function of e.g. magnetization $Z$ ($k=0^+$), current $J$ ($k=0^-$), and staggered magnetization $sZ$ ($k=\pi$). There are characteristic dips around $\tau \approx 0.8$ and $\tau \approx 2.5$, where one therefore expects the strongest chaoticity, i.e., the fastest decay of correlations. Overall, we again see that in $k=0^+$ one has the largest leading RP resonance. Looking only at the $k=0^+$ we see (as was noted already in Ref.~\cite{Prosen,Prosen07}) that there are large section of $\tau$ values where $|\lambda_1|$ appears to be essentially $1$, which would suggest that correlations do not decay, i.e., as if one would have an almost ``integrable'' system (rigorous integrability is of course highly unlikely). Note that we do not have in mind the interval $1.5 \lesssim \tau \lesssim 2.0$ where one indeed has three quite close integrable points at $\tau=\pi/2$ (integrability type (iv) from the introduction), and $\tau=\pi/(2\hx)$ and $\tau=\pi/(2\hz)$ (type (iii) integrability), but regions outside of this interval. However, upon inspecting other sectors we see that things are different. The values of $|\lambda_1|$ there are smaller than $1$, suggesting decay of correlations. We shall explore in more details all those situations for the three indicated selected values of $\tau$ in the following sections. For now let us just say that they correspond to different dynamical regimes, e.g., chaotic with exponential mixing vs. just mixing, and that having access to those other momentum sectors is instrumental in determining the correct dynamical behavior.

Finally, let us comment on the convergence of $\lambda_1$ with $r$. We can see in Fig.~\ref{fig:odtau} that the speed of convergence varies very much with $\tau$ as well as with $k$. For instance, in the $k=0^+$ sector $|\lambda_1|$ has converged already at $r=10$ for almost all shown values of $\tau$ (the red for $r=10$ and the orange curve for $r=8$ are indistinguishable except in a very narrow region around the minimum at $\tau \approx 0.8$). For $k=0^-$ the convergence is good for $\tau>0.65$, but much worse for smaller $\tau$. The situation is least converged for $k=\pi$ (note also that we show only $r=8$), however, there is again a strong asymmetry; the eigenvalue is converged for larger $\tau$, while not at smaller $\tau$. Slow convergence at very small $\tau$ can likely be traced back to the fact that all eigenvalues are very close to $1$, e.g., both the inner and outer radius (\ref{eq:myr}) of a ring that contains most of the eigenvalues converge to $1$ as $\tau \to 0$. For $\tau$ that is an even multiple of $\pi/4$ (the dips in Fig.~\ref{fig:odtau}) the inner radius $r_{\rm in}$ instead becomes $0$, however the number of nonzero eigenvalues of $M(k)$ is smaller and the matrix is strongly non-normal, likely contributing to slow convergence with $r$.

The rich dependence on three relevant parameters calls for a detailed investigation of the KI phase diagram. In the present work we will focus only on three representative cases, showing that there are different dynamical regimes in a quantum many body system. In particular, as we shall see, the system is not always chaotic (see also previous results in Ref.~\cite{Prosen}).

\section{Quantum chaotic regime}

\begin{figure}[t!]
  \centerline{\includegraphics[width=3.in]{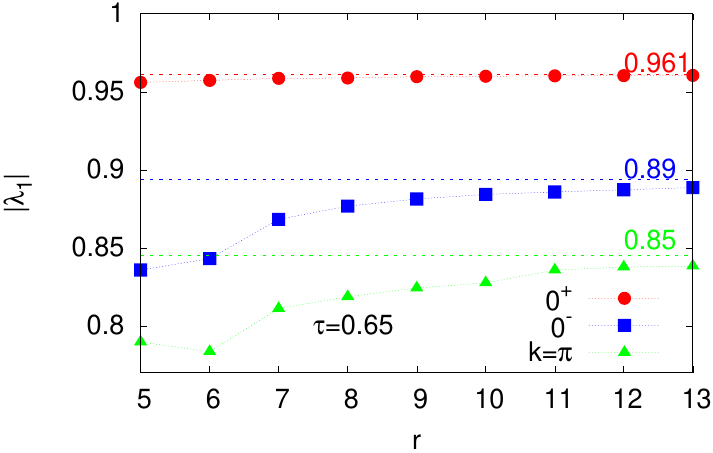}}
\caption{Convergence of the leading RP resonance with $r$ for $\tau=0.65$ and $h_x=0.9$, $h_z=0.8$. The values in the limit $r \to \infty$ are indicated by horizontal dashed lines.}
\label{fig:odr065}
\end{figure}

Let us first focus on $\tau=0.65$ where the leading RP resonance is smaller than $1$ in all $k$ sectors, specifically also for the three $k$ that we focus on, see the middle vertical dashed line in Fig.\ref{fig:odtau}. In Fig.~\ref{fig:odr065} we show more in detail how the leading eigenvalue in each of these sectors changes with increasing $r$, showing that all converge to a value smaller than $1$. The approach to the limiting value seems to be well captured by a finite-$r$ correction scaling as $1/r^2$.

\begin{figure}[ht!]
  \centerline{\includegraphics[width=3.in]{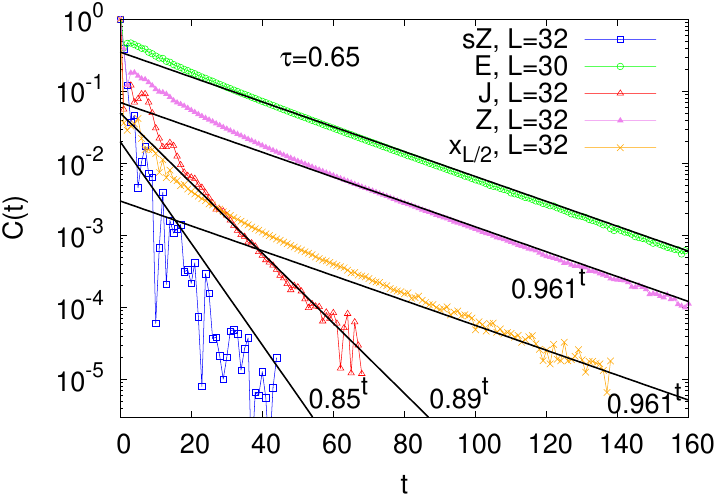}}
\caption{Exponential decay of autocorrelation function $C(t)$ of different observables (see text) in a chaotic regime found at $\tau=0.65$, $\hx=0.9$, $\hz=0.8$. Straight lines are theory based on the calculated RP resonances (Fig.~\ref{fig:odr065}). Correlations are evaluated in a single random initial state, except for the $sZ$ operator where we average over 4.}
\label{fig:065}
\end{figure}

Considering that all $|\lambda_1(k)|<1$ we can say that we are in a quantum chaotic regime. Therefore one expects that all correlation functions should asymptotically decay to zero exponentially as $\sim |\lambda_1|^t$, and, considering that the eigenvalue depends on $k$, different operators will decay with a different rate. We show in Fig.~\ref{fig:065} that this is indeed the case. We compare numerically calculated $C(t)$ and the predicted asymptotic decay based on the leading RP resonance, always finding agreement with values in Fig.~\ref{fig:odr065}.
Observables that are from the $k=0^+$ sector, like the total magnetization $Z$,
\begin{equation}
  Z=\sum_j \sz_j,
  \label{eq:Z}
\end{equation}
or the energy $H_{\rm z}$ (\ref{eq:KI}) denoted by $E$,
\begin{equation}
  E=\sum_j -\sz_j \sz_{j+1}-\frac{\hz}{2}(\sz_j+\sz_{j+1}),
  \label{eq:E}
\end{equation}
decay the slowest. On the other hand, the ``current'' $J$ (\ref{eq:J}) decays faster because it is from the odd $k=0^-$ sector that has smaller $|\lambda_1|$. Still faster is the decay of the staggered magnetization $sZ$ (\ref{eq:sZ}), which is from $k=\pi$ ($sZ$ decays in an oscillatory manner therefore in this case we always plot $|C(t)|$). By superimposing extensive operators with different $k$ we can study operators with any spatial dependence. The extreme case is a strictly local operator, for instance polarization $\sx_{L/2}$ at the middle site, which is a uniform superposition of all $k$ modes. Because the $k=0^+$ contribution will decay the slowest we expect that such a local autocorrelation function will asymptotically also decay as $|\lambda_1(k=0^+)|^t$. The correlation function, defined in this case without an $1/L$ prefactor,
\begin{equation}
  C(t)=\langle \sx_{L/2}(t) \sx_{L/2} \rangle-\langle \sx_{L/2}(t)\rangle \langle \sx_{L/2}\rangle,
\end{equation}
can be seen in Fig.~\ref{fig:065} to indeed decay with the same rate as e.g. the extensive $X$, $Z$, or $E$.

\subsection{Slow convergence with system size}

\begin{figure}[ht!]
  \centerline{\includegraphics[width=3.in]{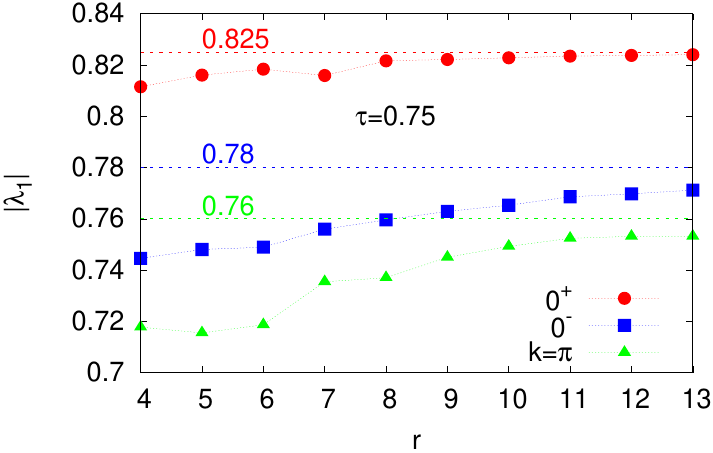}}
\caption{Convergence of leading RP resonances with $r$ for $\tau=0.75$, $h_x=0.9$, $h_z=0.8$.}
\label{fig:odr}
\end{figure}

\begin{figure}[ht!]
  \centerline{\includegraphics[width=3.in]{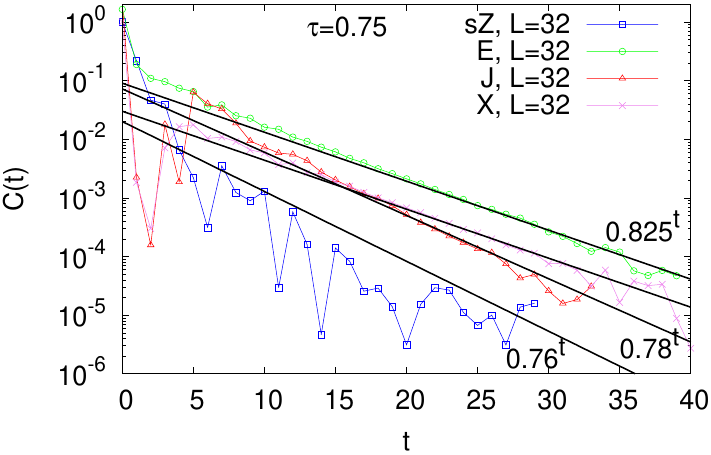}}
\caption{At $\tau=0.75$ and $\hx=0.9, \hz=0.8$ autocorrelation functions of energy, current, magnetization in the x direction, and staggered magnetization all decay with a rate given by the leading RP resonance.}
\label{fig:075}
\end{figure}

\begin{figure}[ht!]
  \centerline{\includegraphics[width=3.in]{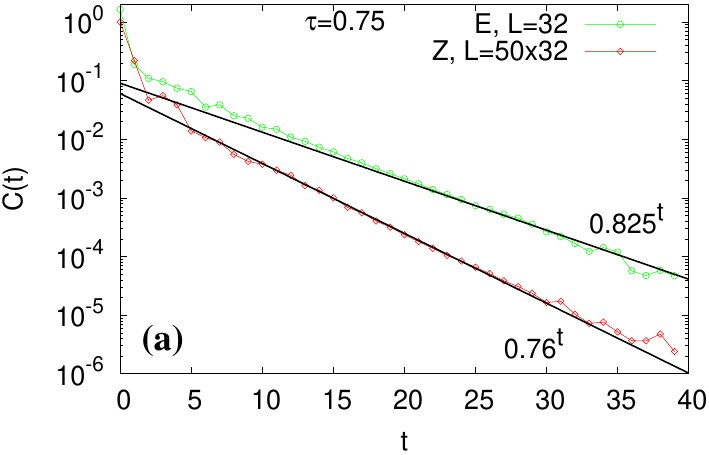}}
    \centerline{\includegraphics[width=3.in]{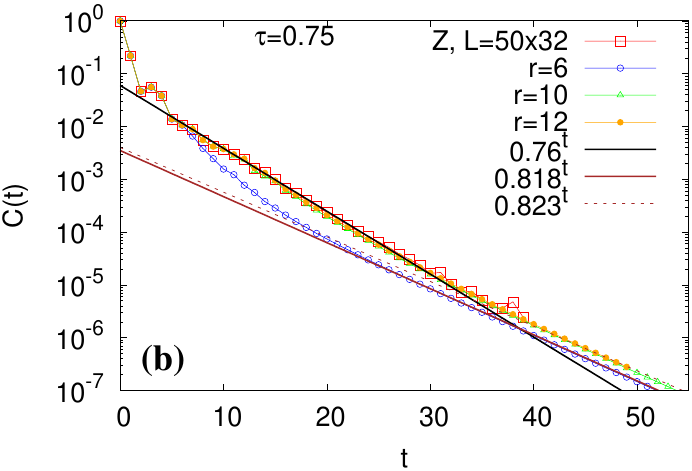}}
    \caption{(a) Magnetization $Z$ (\ref{eq:Z}) in the $z$ direction seems to decay with an ``incorrect'' rate that is larger than $\lambda_1$. (b) Comparison with the evolution by the truncated propagator $M$ ($r=6,10,12$). All is for $\tau=0.75$, $\hx=0.9,\hz=0.8$. Data for $Z$ is averaged over $50$ random initial states with $L=32$. Two straight lines are $\sim 0.818^t$ and $\sim 0.823^t$, corresponding to $\lambda_1$ for $r=6$ and $r=10$ ($\lambda_1$ for $r=12$ rounded to three digits is $0.824$).}
\label{fig:075Z}
\end{figure}

Next we study $\tau=0.75$, which, according to the rightmost vertical dashed line in Fig.~\ref{fig:odtau}, is even more strongly chaotic with smaller values of the leading RP resonances that again converge well with $r$, see Fig.~\ref{fig:odr}. 
As visible in Fig.~\ref{fig:075} the energy $E$ (\ref{eq:E}), the total magnetization $X$,
\begin{equation}
  X=\sum_j \sx_j,
  \label{eq:X}
\end{equation}
the staggered magnetization $sZ$ (\ref{eq:sZ}), and the current $J$ (\ref{eq:J}), all decay with the expected rate given by the respective RP resonance.
However, looking at the magnetization $Z$ (\ref{eq:Z}) in Fig.~\ref{fig:075Z}(a) an exponential decay is seen with a rate that is not given by $|\lambda_1(0^+)|$. This is rather puzzling considering that the RP resonance $\lambda_1(0^+)$ converges rather quickly (Fig.~\ref{fig:odr}), and that this seeming incorrect decay persists over almost 3 decades all the way till finite-size fluctuations become noticeable at very small values of $C(t)$. To understand how that can be we have looked at the evolution of $Z$ by the truncated propagator $M(0^+)$ whose largest eigenvalue is clearly $\lambda_1 \approx 0.825$ (and is different from the observed decay that goes as $\sim 0.76^t$). Namely, within the truncation to $r$-site operators the corresponding correlation function can be simply written as
\begin{equation}
  C(t) \approx \mathbf{y}^T\cdot  M(0^+)^t\cdot \mathbf{y},
  \label{eq:CM}
\end{equation}
where the initial vector $\mathbf{y}$ corresponds to the operator $a \in \Sr$, in our case to $a=\sz$. In a basis where $a=\sz$ is chosen as one basis vector one therefore needs only the appropriate diagonal matrix element of the truncated matrix $M^t$. 
In Fig.~\ref{fig:075Z}(b) we can see that the above truncation approximation to $C(t)$ (Eq.\ref{eq:CM}, curves with small symbols in Fig.~\ref{fig:075Z}(b)) does asymptotically decay as $\lambda_1^t$, but the asymptotic decay starts rather late, say around $t>40$ when $C(t)$ is already quite small (because the corresponding expansion coefficient is small). We can also see that, while the $r=6$ approximation does not yet capture the true decay for $t<30$, higher truncations like $r=10$ or $r=12$ do. We can conclude that in order to see the true asymptotic decay $|\lambda_1|^t$ one needs to look at small values $C(t)< 10^{-6}$, and therefore in order to even get a glimpse of it in a single random state evolution one would need large systems with $L>40$.

Such decay, where one initially has faster relaxation $\sim 0.76^t$ that then eventually goes into asymptotic $\sim 0.82^t$ given by the largest eigenvalue, is reminiscent of a recently discovered two-step relaxation in random circuits~\cite{PRX} in which the initial decay is determined by the pseudospectrum rather than the spectrum~\cite{U4}. It can happen when dealing with non-normal matrices, e.g., non-Hermitian, that are badly conditioned. In such cases the spectrum is a rather singular object that is very sensitive to perturbations while the pseudospectrum~\cite{trefethen} is robust. It seems though that the effect here is different in one crucial aspect: the window in which one has the ``incorrect'' decay $\sim 0.76^t$ is here of fixed width whereas in the mentioned random circuits it diverges with system size. This divergence comes because the condition number of those matrices in random circuits diverges with system size, allowing for an increasing window size as one increases the system size. To get an idea on the condition number of our $M(k=0^+)$ we have looked at its singular values. Namely, for a nonsingular matrix $M$ its condition number is simply $\kappa=s_{\rm max}(M)/s_{\rm min}(M)$, where $s_{\rm max}$ and $s_{\rm min}$ are maximal and minimal singular values. Using exact expressions for singular values (\ref{eq:s}) we see that the the maximal one is $1$, while the minimal one is $\cos^2{(2\tau)}$, both independent of $r$ or any other parameter. The matrix $M$ becomes singular only at special $\tau=\pi/4\approx 0.78$. Because our $\tau=0.75$ is very close to that singularity we have a rather large $\kappa \approx 200$, which though is finite and in particular does not increase with $r$. Therefore, there is no true two-step relaxation; the initial decay is just a transient that happens to hold for a rather large range of $C(t)$. Nevertheless, slow convergence of the autocorrelation function towards its asymptotic decay can be traced back to strong non-normality of $M(k)$. An interesting open question is, whether perhaps at $\tau=\pi/4$ one could have some effects of strong non-normality due to some singular values becoming zero.
\begin{figure}[t!]
  \centerline{\includegraphics[width=3.2in]{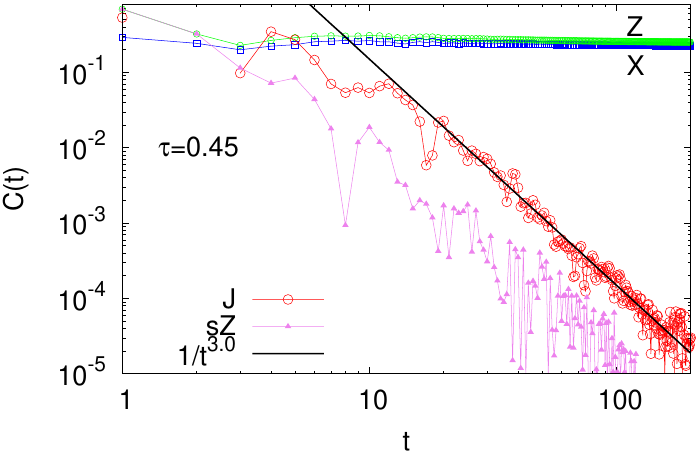}}
\caption{Autocorrelation functions from different sectors for $\tau=0.45$ and $\hx=0.9, \hz=0.8$. Straight line suggests a power-law decay $\sim 1/t^{3.0}$. One random initial state is used, for $X$ and $Z$ with $L=30$, for $sZ$ with $L=32$, and for $J$ size is $L=33$.}
\label{fig:045}
\end{figure}

\section{Mixing regime}

The last parameter set that we look at is $\tau=0.45$ and $\hx=0.9, \hz=0.8$ (the leftmost vertical dashed line in Fig.~\ref{fig:odtau}). This regime looks special because while $|\lambda_1(k=0^+)|\approx 1$ the largest eigenvalues in other sectors look distinctively smaller than $1$ at finite $r$.

\subsection{Power-law decay}

Looking at numerically computed autocorrelation functions, Fig.~\ref{fig:045}, we see that indeed the ones from the $k=0^+$ do not seem to decay. This looks similar to what was called (for other parameter values) an intermediate regime in Ref.~\cite{Prosen07}. On the other hand, inspecting other sectors, for instance the current $J$ (\ref{eq:J}), which is from $k=0^-$, or the staggered magnetization $sZ$ (\ref{eq:sZ}) which is from $k=\pi$, we can see a clear decay. Decay in fact looks like a power-law $\sim 1/t^\alpha$ with a power $\alpha \approx 3$ over more than $2$ decades in $C(t)$.

\begin{figure}[t!]
  \centerline{\includegraphics[width=3.2in]{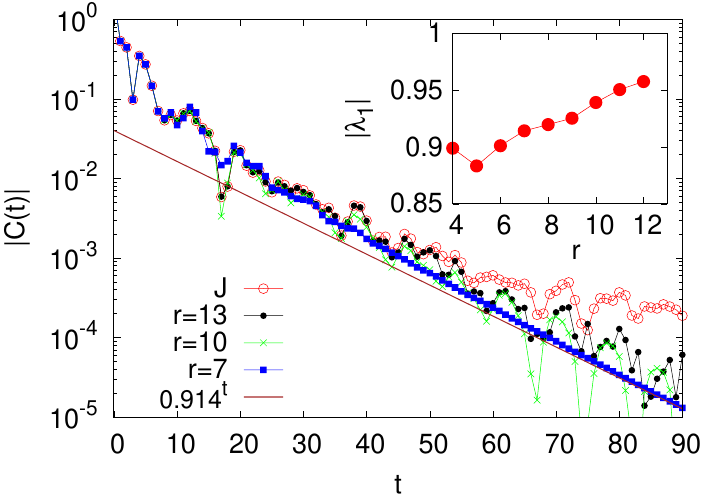}}
  \caption{Exact correlation function of $J$ and $L=33$ (red circles, the same as in Fig.\ref{fig:045}), and approximations by $M(0^-)$ (\ref{eq:CM}) for different $r$. Brown line is asymptotics with $\lambda_1$ for $r=7$. The inset shows the largest eigenvalue as a function of $r$. All is for $\tau=0.45, \hx=0.9, \hz=0.8$.}
\label{fig:045_conv}
\end{figure}

\begin{figure}[t!]
  \centerline{\includegraphics[width=2.7in]{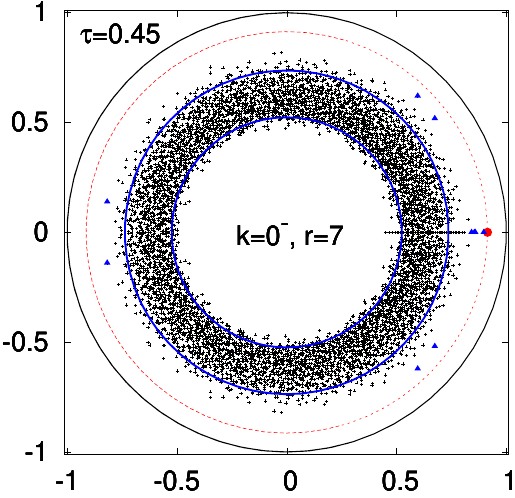}}
\caption{Eigenvalue spectrum of $M(k=0^-)$ for $\tau=0.45, \hx=0.9, \hz=0.8$ and $r=7$. There are $5985$ eigenvalues, with the largest at $\lambda_1\approx 0.914$ being highlighted by a red circle, with further 10 with the largest modulus being shown with small blue triangles (three largest ones are $\lambda_2 \approx 0.897$, and a complex pair $|\lambda_{3,4}|\approx 0.861$). Blue circles are theoretical $r_{\rm in}$ and $r_{\rm out}$ (\ref{eq:myr}).}
\label{fig:045spek}
\end{figure}

To get a better understanding of how the power-law decay arises we have looked more closely at how the approximation of exact $C(t)$ with a finite-$r$ matrix $M(k)$ (\ref{eq:CM}) works. Results are shown in Fig.~\ref{fig:045_conv}.
We can see that at each finite $r$ the asymptotic decay of $\mathbf{y}^T M \mathbf{y}$ is actually exponential, in-line with $|\lambda_1|<1$, but the rate of this exponential decay decreases with increasing $r$ (the inset). Increasing $r$ the approximation with $M(0^-)$ gets better, but the convergence towards numerically computed power-law decay of $C(t)$ for $L=33$ is rather slow. With increasing $r$ the largest eigenvalue $|\lambda_1(k=0^-)|$ will converge towards $1$. In Fig.~\ref{fig:045spek} we show an example of a full spectrum of $M(k=0^-)$ that suggests what is going on: in addition to an RMT-like annular cloud of eigenvalues, whose boundaries are well described by Eq.(\ref{eq:myr}), we can also see a large number of eigenvalues on the real axis. That is similar to what has been recently seen~\cite{arul24} in channels describing reduced dynamics of coupled standard maps in a regime of no KAM tori but with a stable fixed point. We conjecture that as one increases $r$ the number of real eigenvalues increases, pushing the largest one $|\lambda_1|$ towards $1$, while the resulting continuum of eigenvalues, i.e., a branch cut, is responsible for the power-law decay. Note that while this scenario is not yet very clear in the shown spectrum for $r=7$, the power-law decay is quite clear in a directly calculated $C(t)$ for $L=33$.

\subsection{Prethermalization in $k=0^+$}

\begin{figure}[t!]
  \centerline{\includegraphics[width=2.8in]{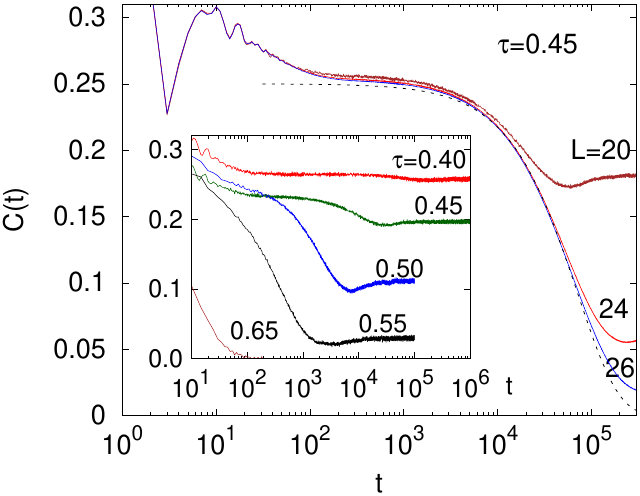}}
\caption{Autocorrelation function of magnetization $Z$ (\ref{eq:Z}) at $\hx=0.9, \hz=0.8$. Main plot shows convergence with $L$ at $\tau=0.45$ (dashed black curve is $0.25\cdot \lambda_1^t$ with $\lambda_1=1-14\cdot 10^{-6})$, while the inset shows dependence on $\tau$ at fixed system size $L=18$.}
\label{fig:045Z}
\end{figure}
Considering the above explanation we must also rethink the apparent plateau seen in the $k=0^+$ sector (Fig.~\ref{fig:045}). Namely, it seems more natural that one would have a decay in all sectors. And indeed, looking at the autocorrelation function of $Z$ (\ref{eq:Z}) at much longer times we eventually do see a decay, Fig.~\ref{fig:045Z}. At a finite $L$ there appears to be a 2nd plateau at a late time, e.g. $t>10^5$ at $L=20$, which though, as seen in the figure, is just a finite-size effect. As one increases $L$ one approaches the asymptotic exponential decay given by the largest eigenvalue that is very close to $1$, namely $\lambda_1\approx 1-14\cdot 10^{-6}$ (for studied truncations $\lambda_1$ seems rather stable with only small even-odd $r$ oscillations). This exponential decay only starts after very long transient 1st plateau of duration $\sim 10^3$. Looking at the $\tau$ dependence in the inset of Fig.~\ref{fig:045Z} we see that the time and the height of this 2nd finite-size plateau rapidly change with $\tau$. We therefore have a rather interesting situation: observables from the $k=0^-$ sector decay (quickly) as a power law, while those from the $k=0^+$ eventually do decay exponentially but only after a long plateau.

\begin{figure}[t!]
  \centerline{\includegraphics[width=1.67in]{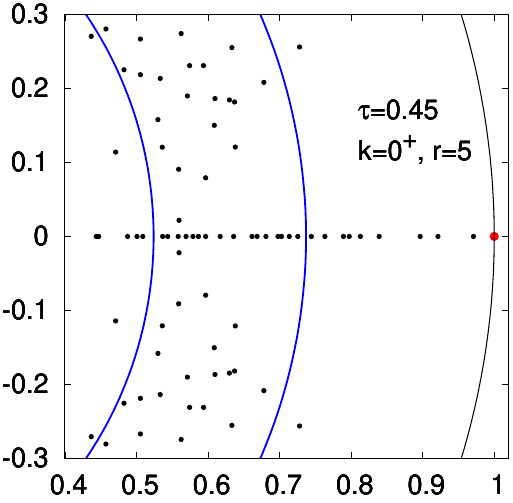}\hskip2pt\includegraphics[width=1.67in]{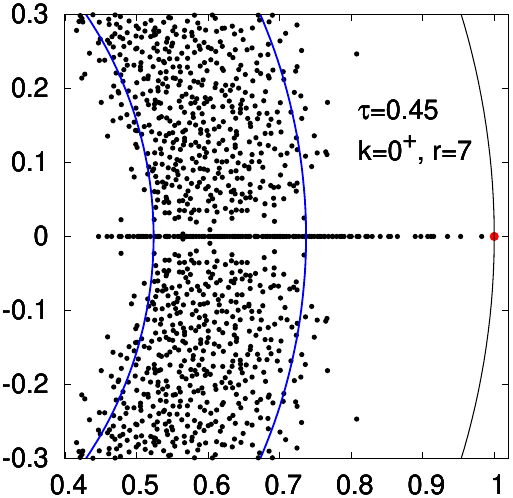}}
  \vskip2pt
  \centerline{\includegraphics[width=2.7in]{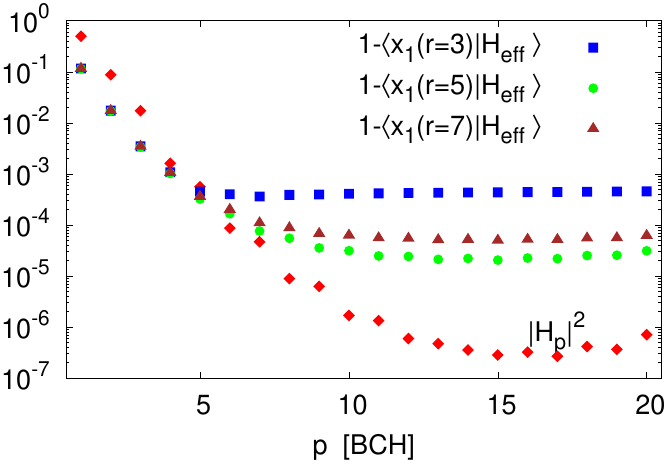}}
  \caption{An almost conserved operator in the mixing regime in sector $k=0^+$ for $\tau=0.45,\, \hx=0.9, \hz=0.8$. Top: Zoom-in on the spectrum around the largest $\lambda_1\approx 1-14\cdot 10^{-6}$ highlighted in red (blue circles are theoretical $r_{\rm in}$ and $r_{\rm out}$), for $r=5$ (left, 423 eigenvalues), and $r=7$ (right, 6303 eigenvalues). Bottom: The eigenvector $\mathbf{x}_1$ corresponding to $\lambda_1$ is for small $r$ equal to $H_{\rm eff}$ obtained by summing first $p$ BCH orders. Norm of the BCH term $H_p$ of order $p$ is shown with red diamonds.}
\label{fig:045spekS}
\end{figure}
The long plateau can be explained by what has been called a prethermalization~\cite{keiji16}. Namely, although $\tau$ does not necessarily look perturbatively small one can nevertheless try to write the effective one-step Floquet generator $H_{\rm eff}$ using the Baker-Campbell-Hausdorff (BCH) expansion
\begin{eqnarray}
  U={\rm e}^{H_{\rm eff}},\quad H_{\rm eff}&=&-\ii \tau (H_{\rm x}+H_{\rm z}) -\frac{\tau^2}{2} [H_{\rm x},H_{\rm z}]+\cdots \nonumber \\
  &=&\sum_{p=1} H_p,
\end{eqnarray}
with $H_p$ being the antihermitian $p$-th order term (in $\tau$) in the BCH expansion. If this expansion would converge, and therefore could describe the long-time dynamics by simple $U^t={\rm e}^{H_{\rm eff}t}$, the $H_{\rm eff}$ would be a constant of motion. Provided it would be a local operator this should be reflected in the spectral properties of $M(k=0^+)$, i.e., one eigenvalue being $1$. However, the formal BCH expansion is usually divergent with terms $H_p$ eventually growing for $p>p_0$. Up until times when divergent $H_{p>p_0}$ become important the $H_{\rm eff}$ truncated at the order $p_0$ is almost conserved~\cite{keiji16}, i.e., its exact lack of conservaton will become noticeable only at large times. Such an almost conserved $H_{\rm eff}$, a hallmark of prethermalization~\cite{keiji16}, will result in one eigenvalue of $M(k=0^+)$ being very close to $1$. And this is exactly what happens in our kicked Ising model at $\tau=0.45$. This can be seen in Fig.~\ref{fig:045spekS}. As already mentioned, the largest eigenvalue $\lambda_1$ is very close to $1$ with the corresponding eigenvector $\mathbf{x}_1$ being, upto normalization, exactly equal to the truncated $H_{\rm eff}$. This is verified in the bottom frame of Fig.~\ref{fig:045spekS} where we calculate the overlaps between $\mathbf{x}_1$ calculated from $M(0^+)$ for a finite support size $r$, say $r=3$, and compare it with $H_{\rm eff}$ obtained by summing up the first $p$ orders of the BCH series. Due to the structure of the kicked Ising model the support of the BCH term $H_p$ of order $p$ is $\sim 1+p/2$, i.e., every 2nd order increases the support of commutators by $1$ site. This can be nicely seen from the data: for $r=3$ the fidelity saturates at $p\approx 6$, while for $r=5$ at around $p\approx 10$. One can also see that the minimal error of $\mathbf{x}_1$ for a given $r$, i.e., its difference from $H_{\rm eff}$, is essentially equal to the norm $|H_p|^2:={\rm tr}(H_p^\dagger H_p)$ of the truncated terms (red diamonds in the figure) for $p \approx 2r$, provided we are in a convergent regime $p<p_0$, that is, as long as $H_p$ decrease. Namely, by calculating BCH terms upto order $20$~\cite{casas08} we can see (Fig.~\ref{fig:045spekS}) that the terms indeed start to increase after $p_0 \approx 16$. This means that $H_{\rm eff}$ is not really conserved and we do not expect any eigenvalue of $M(k=0^+)$ to be exactly $1$. The eventual divergent nature of the BCH series is also reflected in the structure of the eigenvector $\mathbf{x}_1$ -- for larger $r$ it ceases to be well approximated by $H_{\rm eff}$ (compare $r=5$ and $r=7$ in the figure) -- as well as in the spectrum of $M(0^+)$ (other eigenvalues moving closer to $\lambda_1$).

One question that remains is what happens in the thermodynamic limit $L \to \infty$, in particular in view of a seeming power-law decay in other sectors? There are two possibilities. As we can see in the top Fig.~\ref{fig:045spekS} there are many eigenvalues on the real axis, the number of which increases with $r$, similarly as is the case in the $k=0^-$ sector. So it could be that at very large $r$ when those eigenvalues move really close to $1$ they begin to mix with $\lambda_1$ which eventually loses its ``prethermalization'' character of describing $H_{\rm eff}$, altogether resulting in a power-law decay of correlations also in the $k=0^+$ sector. Another, perhaps less likely scenario would instead be that the decays in both $k=0^+$ and $k=0^-$ stay as they are in figures, an exponential and a power-law, respectively. In such a case the decay in the $k=0^+$ sector would eventually become faster than the one in $k=0^-$ (exponential decay always wins over a power law), meaning that the resonance in $k=0^+$ is not always the largest. Note that due to prefactors ($0.25\cdot 0.999986^t$ vs. $150/t^3$) $C(t)$ for the current $J$ would be larger than the one for $Z$ only for $t > 2.5\cdot 10^6$ when $C(t) < 10^{-17}$ (requiring $L>120$ to see it in a single random state).

We find it quite interesting that one can extract all prethermalization information from the truncated operator propagator $M(k)$ -- a procedure that seems simpler than for instance tediously evaluating BCH terms to a high order. It would be also interesting to understand the physical mechanism behind this slow relaxation and whether behavior in different momentum sectors are somehow related. A power-law decay of correlations at a finite $\tau=0.45$ means that in the thermodynamic limit and for finite breaking of integrability one can have situations without chaos with exponential decay of correlations. Such slow breaking of integrability is a bit akin to what one expects in classical systems based on the KAM theorem~\cite{ott}. Upon weak perturbation some classical orbits (KAM torii) remain essentially integrable, e.g., they display non-decaying autocorrelation functions. However, in a high dimensional system they do not separate the phase space into an inside and outside and orbits can escape (Arnold diffusion).

\section{Conclusion}

We studied momentum dependent Ruelle-Pollicott resonances of a truncated operator propagator in translationally invariant quantum systems, focusing on the kicked Ising model. They give much better insight into dynamics than looking only at translationally invariant observables, i.e., from the zero momentum subspace. We always find that the leading resonance of the truncated propagator correctly predicts the asymptotic decay of correlation functions.

While we essentially always find that the largest resonance is from the even $k=0$ sector, it is not clear if there is a proof that this always has to be the case. In fact, a counterexample could be a very interesting parameter regime where it seems that some correlations decay as a power-law due to a branch cut, while other decay exponentially but with a very small decay rate. This separation of timescales can be explained in terms of an almost conserved operator (prethermalization) that can be alternatively calculated via a BCH formula. While it is believed that the kicked Ising model is a generic chaotic toy model we see that its dynamics has a rich dependence on parameters. To that end it would be interesting to study Ruelle-Pollicott resonances in other systems, for instance without a commuting interacting part, including in generic quantum circuits. 

Based on the conjectured exact expression for singular values of the truncated propagator we propose a uniform lower bound on resonances. Better understanding properties of the truncated propagator, especially at $\tau=\pi/4$ (which includes a dual unitary case), is also an open problem.

\section*{Acknowledgments}
I would like to thank Urban Duh for cooperation on related projects, and Arul Lakshminarayan, Lucas S\' a, Toma\v z Prosen, and Takashi Mori for discussions and comments. I also acknowledge support by Grants No.~J1-4385 and No.~P1-0402 from Slovenian Research Agency.

\appendix

\section{Different parametrizations}
\label{app:Par}

A product of any number of rotations can always be written as a single rotation by some angle around a direction specified by a unit vector $\mathbf{n}$. Specifically, having a rotation around the z axis followed by a rotation around the x axis, like in our parameterization of the KI model, we can write
\begin{equation}
  {\rm e}^{\ii \alpha \sx} {\rm e}^{\ii \beta \sz}={\rm e}^{\ii \gamma \mathbf{n}\cdot \boldsymbol{\sigma}},
  \label{eq:rot} 
\end{equation}
where the strength of the field is given by $\cos{\gamma}=\cos{\alpha}\cos{\beta}$, while its direction is $\mathbf{n}=(\sin{\alpha}\cos{\beta},\sin{\alpha}\sin{\beta},\cos{\alpha}\sin{\beta})/\sin{\gamma}$. Because the interaction is invariant to rotations around the z axis, we can rotate the propagator around the z axis in order to have the field on the RHS of Eq.(\ref{eq:rot}) only in the x-z plane. For the above $\mathbf{n}$ this is achieved by rotation $V={\rm e}^{-\ii \beta \sz/2}$, resulting in
\begin{equation}
V^\dagger {\rm e}^{\ii \gamma \mathbf{n}\cdot \boldsymbol{\sigma}} V={\rm e}^{\ii \gamma \mathbf{n}' \cdot \boldsymbol{\sigma}}, 
\end{equation}
with
\begin{equation}
  \mathbf{n}'=(\sin{\theta},0,\cos{\theta}), \quad \tan{\theta}=\frac{\tan{\alpha}}{\sin{\beta}},\quad \cos{\gamma}=\cos{\alpha}\cos{\beta}.
  \label{eq:gamma}
\end{equation}
As an example, parameters $\tau=0.65$, $\hx=0.9$ and $\hz=0.8$ are equivalent to a KI model with $U={\rm e}^{\ii J \sum_j \sz_j \sz_{j+1}} {\rm e}^{\ii \sum_j \hx' \sx_j+\hz' \sz_j}$ (like the one studied in Ref.\cite{Prosen}) and $J=0.65$ and $\hx'\approx 0.61$, $\hz'\approx 0.46$.

\section{Matrices}
\label{app:matrix}

The propagator of the KI model (\ref{eq:KIU}) can be decomposed into a series of one and two-qubit gates. To evaluate transformation of a local operator (\ref{eq:ca}) (and therefore also the matrix elements of $M(k)$, if needed) one applies a series of those gates to a $4^{r+2}$ dimensional vector encoding the local operator on $r+2$ sites. Transformations induced on the space of Pauli matrices by the gates of the KI model are rather simple.

All matrices needed can in fact be written in terms of a $2\times 2$ matrix
\begin{equation}
  P(a)=
  \begin{pmatrix}
	\cos{(2\tau)} & -\sin{(2\tau)}\\
	\sin{(2\tau)} & \cos{(2\tau)}
  \end{pmatrix}.
  \label{eq:P}
\end{equation}
The rotation around the x-axis given by $U_{\rm x}={\rm e}^{\ii \tau \sx}$ induces via $a'=U_{\rm x}^\dagger a U_{\rm x}$ the following transformation
\begin{equation}
  O_{\rm x}=
  \begin{pmatrix}
	1  & 0&0  &0 \\
	0 & \cos{(2\tau)} & -\sin{(2\tau)} &0\\
	0 & \sin{(2\tau)} & \cos{(2\tau)} &0\\
        0 & 0 & 0 & 1\\
	\end{pmatrix}=P(\tau)_{y,z} \oplus\, \1_{x,\1},
\end{equation}
where the operator basis is ordered as $\sx,\sy,\sz,\1$, and where in a subscript we list the set of basis states on which the matrix acts.

Rotation around the z-axis is similar,
\begin{equation}
  O_{\rm z}=P(\tau)_{x,y} \oplus \1_{z,\1}.
\end{equation}

While the 2-site gate $U_{\rm zz}={\rm e}^{\ii \tau \sz_1 \sz_2}$ describing interaction is
\begin{eqnarray}
  O_{\rm zz}=&&P(\tau)_{zx,1y} \oplus P(\tau)_{xz,y1} \oplus P(\tau)_{1x,zy} \oplus P(\tau)_{x1,yz} \oplus \nonumber \\
  &&\oplus \1_{xx,yx,xy,yy,zz,1z,z1,11}.
\end{eqnarray}
At $\tau=\pi/4$ it is a signed permutation matrix, i.e., a permutation matrix with all diagonal matrix elements being zero, while half of the off-diagonal ones is $-1$ (it is an antisymmetric matrix).

The propagation of operators by the whole one-step $U$ is then performed by applying first the gate $O_{\rm zz}$ to all n.n. pairs, then $O_{\rm z}$ to all sites, and finally $O_{\rm x}$ to all sites.


\begin{thebibliography}{99}
\bibitem{foot5} As one deals with non-unitary operators it is at present not clear if there are cases where one would have to in fact look at pseudoeigenvalues~\cite{trefethen} instead of at eigenvalues, as was the case is some recent studies of relaxation, e.g. Refs.~\cite{U4}.
\bibitem{foot3}  We note that if one would have a truncated Haar random unitary matrix eigenvalues would lie within a disk, i.e., the inner radius would be zero~\cite{karol00}.
\bibitem{foot2} The only possible exception could be the dual unitary case of $\tau=\pi/4$ and $\hx=1$, which is clearly very special, and for which it is not clear if the isolated eigenvalues of $M(k)$ play the role of traditional RP resonances. For instance, at $\hz=1$ all eigenvalues of $M(k)$ are zero and one has a Clifford circuit -- a circuit that does not mix different product of Pauli matrices and whose dynamics has only polynomial complexity (i.e., a Pauli product is mapped to some other single Pauli product).
\bibitem{foot1} For larger $r$ we use a simple power method to get the eigenvalue with the largest modulus, or the ARPACK package to get few eigenvalues with the largest $|\lambda_j|$.
\end{thebibliography}
\end{document}